\documentclass[preprint,showpacs,preprintnumbers,amsmath,amssymb,prb,aps]{revtex4-1}
\usepackage{epsfig}% Include figure files
\usepackage{epstopdf}
\usepackage{graphicx}% Include figure files
\usepackage{dcolumn}% Align table columns on decimal point
\usepackage{bm}% bold math
\usepackage{slashbox}
\usepackage{textcomp}
\begin{document}

\title{Constrained DFT+$U$ approach for understanding the magnetic behaviour of ACr$_{2}$O$_{4}$ (A=Zn, Mg, Cd and Hg) compounds}
\author{Sohan Lal and Sudhir K. Pandey}
\affiliation{School of Engineering, Indian Institute of Technology Mandi, Kamand 175005, Himachal Pradesh, India}

\date{\today}

\maketitle

\section{Abstract}

In this work, we try to understand the inconsistency reported by [Yaresko, Phys. Rev. B. {\bf 77}, 115106 (2008)] in the theoretically estimated nature and the variation of magnitude of nearest neighbour exchange coupling constant ($\arrowvert${\it J$_{1}$}$\arrowvert$) with increasing $U$ in ACr$_{2}$O$_{4}$ (A=Zn, Cd, Mg and Hg) compounds by using density functional theory. In unconstrained calculations, the nature and variation of $\arrowvert${\it J$_{1}$}$\arrowvert$ as a function of $U$ in the present study are not consistent with the experimental data and not according to the relation, {\it J$_{1}$}$\propto$$\frac{t^{2}}{U}$ especially for CdCr$_{2}$O$_{4}$ and HgCr$_{2}$O$_{4}$ for $U >$3 eV and U=2-6 eV, respectively. Such an inconsistent behavior of $\arrowvert${\it J$_{1}$}$\arrowvert$ is almost similar to that of Yaresko for these two compounds for $U$=2-4 eV. For ZnCr$_{2}$O$_{4}$ and MgCr$_{2}$O$_{4}$, the nature and the variation of $\arrowvert${\it J$_{1}$}$\arrowvert$ in the present work are in accordance with the experimental data and above mentioned relation for $U$=2-6 eV and are similar to that of Yaresko for ZnCr$_{2}$O$_{4}$ for $U$=2-4 eV. However, in constrained calculations the nature and variation of $\arrowvert${\it J$_{1}$}$_{c}$$\arrowvert$ in the present work are according to experimental data and above above mentioned relation for all four compounds. Hence, the present study shows the importance of constrained calculations in understanding the magnetic behaviour of these spinels. The values of magnitude of Curie-Weiss temperature [$\arrowvert$($\varTheta$$_{CW}$)$_{c}$$\arrowvert$] for ZnCr$_{2}$O$_{4}$$>$MgCr$_{2}$O$_{4}$$>$CdCr$_{2}$O$_{4}$$>$HgCr$_{2}$O$_{4}$ for $U$=2-5 eV, which are according to the order of experimentally observed values for these spinels. The calculated values of ($\varTheta$$_{CW}$)$_{c}$ for ZnCr$_{2}$O$_{4}$, MgCr$_{2}$O$_{4}$, CdCr$_{2}$O$_{4}$ and HgCr$_{2}$O$_{4}$ are -982 K, -721 K, -147 K and -122 K, respectively at $U$=5 eV.

\section{Introduction}

 It is well known that the magnetic structure and properties of magnetic materials (eg. Curie-Weiss temperature ($\varTheta$$_{CW}$)) are mainly determined by the magnitude and the sign of the interatomic exchange coupling constants ($J_{ij}$) arising among the magnetic ions. Because of the sufficiently high $\varTheta$$_{CW}$ (above room temperature), any magnetic material can be used in the new generation electronic devices.\cite{Parkin} Hence, the ability of first principles electronic structure calculations to predict the $J_{ij}$ plays an important role for designing of the various materials. Theoretically, the basic mechanisms to predict the $J_{ij}$ are quite well-known.\cite{Fazekas} However, it is not easy task to predict the sign and magnitude of $J_{ij}$ parameters for real magnetic materials. Liechtenstein $et$ $al$. proposed a general method to extract the exchange integrals from electronic structure calculations, where the energy of the electronic Hamiltonian is mapped onto a classical Heisenberg model.\cite{Liechtenstein1987} Nowadays, this approach has become a prominent theoretical tool for studying the inter-site magnetic interactions for different materials from $ab$ $initio$ electronic structure calculations.\cite{Wysocki,Mazurenko}

     Transition metal spinel oxides, ACr$_{2}$O$_{4}$ [where A=(Zn, Mg, Cd, Hg) and Cr are non-magnetic and magnetic sites, respectively] are complex systems, where the inter-site magnetic interactions have been extensively studied from about last two decades.\cite{Martinho,Chung,Fennie,Chan,Yaresko,Xiang,Kumar,Kimura} At room temperature, these spinels have face centered cubic {\it Fd$\bar{3}$m} crystal structure, where the magnetic Cr ions form a pyrochlore lattice.\cite{Tchernyshyov88,Matsuda2007} A strong geometrical frustration arises in these spinels due to the three-dimensional network of corner sharing tetrahedra of antiferromagnetically coupled Cr ions.\cite{Tchernyshyov88} In these compounds, the Cr$^{3+}$ ion has three electrons in t$_{2g}$ orbitals. The direct overlap of these orbitals of neighboring Cr sites give the dominant antiferromagnetic (AFM) nearest-neighbor interactions in the spin Hamiltonian. The presence of strong geometrical frustration, these compounds remain in paramagnetic phase well below the $\varTheta$$_{CW}$. Experimentally observed values of $\varTheta$$_{CW}$ for ZnCr$_{2}$O$_{4}$, MgCr$_{2}$O$_{4}$, CdCr$_{2}$O$_{4}$ and HgCr$_{2}$O$_{4}$ are ($\sim$-392 K, $\sim$-398 K), ($\sim$-350 K, $\sim$-346 K), ($\sim$-83 K, $\sim$-71 K) and $\sim$-32 K, respectively.\cite{Baltzer,Ueda2007,Rudolf} However, in some literatures the values of $\varTheta$$_{CW}$ are reported to be $\sim$-400 K, $\sim$-433 K and $\sim$-70 K for ZnCr$_{2}$O$_{4}$, MgCr$_{2}$O$_{4}$ and CdCr$_{2}$O$_{4}$, respectively, which indicate the dependence of $\varTheta$$_{CW}$ on the fitting of reciprocal of magnetic susceptibility versus temperature data.\cite{Takagi,Dutton} The negative sign of $\varTheta$$_{CW}$ for these compounds indicates the AFM nature of the ground state. However, the exact AFM spin structure for these compounds are unknown.\cite{Rudolf,Dutton,Martin,Lee,Chung2013}

   Theoretically, Yaresko has calculated the values of $\varTheta$$_{CW}$ for ZnCr$_{2}$O$_{4}$, CdCr$_{2}$O$_{4}$ and HgCr$_{2}$O$_{4}$ compounds by using the calculated values of exchange coupling constants ($J_{n}$) up to the fourth Cr neighbours for $U$=2 to 4 eV. In this range of $U$, the values of $J_{n}$ are obtained by fitting the energy of spin spirals to a classical Heisenberg model, where the sign of the magnetic energy is considered to be +ve as a convention. According to  Yaresko, +ve and -ve values of $J_{n}$ corresponds to AFM and FM interactions, respectively. The sign of $\varTheta$$_{CW}$ is mainly determined by the first nearest neighbour ($J_{1}$) because of its dominant contribution as compared to second ($J_{2}$), third ($J_{3}$) and fourth ($J_{4}$) nearest neighbours. The values of $\varTheta$$_{CW}$ for ZnCr$_{2}$O$_{4}$ changes from -500 K to -328 K to -209 K when $U$ changes from 2 to 3 to 4 eV, which are consistent with the experimentally reported sign of $\varTheta$$_{CW}$. In this range of $U$, negative sign of $\varTheta$$_{CW}$ for ZnCr$_{2}$O$_{4}$ are mainly due to the dominant AFM $J_{1}$, which changes from $\sim$5.26 to $\sim$3.45 to $\sim$2.15 meV as $U$ varies from 2 to 3 to 4 eV as per expectation, if $J$$\propto$$\frac{t^{2}}{U}$. For CdCr$_{2}$O$_{4}$, $\varTheta$$_{CW}$ ($J_{1}$) changes from -64 K ($\sim$0.52 meV) to 12 K ($\sim$-0.34 meV) to 62 K ($\sim$-0.77 meV) as $U$ varies from 2 to 3 to 4 eV. Similarly for HgCr$_{2}$O$_{4}$, the values of $\varTheta$$_{CW}$ ($J_{1}$) changes from -14 K ($\sim$-0.60 meV) to 59 K ($\sim$-1.21 meV) to 104 K ($\sim$-1.55 meV) when $U$ varies from 2 to 3 to 4 eV. Estimated sign of $\varTheta$$_{CW}$ as well as $J_{1}$ for CdCr$_{2}$O$_{4}$ and HgCr$_{2}$O$_{4}$ are inconsistent with experimentally reported sign of $\varTheta$$_{CW}$ as well as not according to the equation, $J$$\propto$$\frac{t^{2}}{U}$ for $U >$2 eV. According to Yaresko, as the lattice constant increases from ZnCr$_{2}$O$_{4}$ to CdCr$_{2}$O$_{4}$ to HgCr$_{2}$O$_{4}$, the strength of the direct $d$-$d$ hopping decreases from ZnCr$_{2}$O$_{4}$ to CdCr$_{2}$O$_{4}$ to HgCr$_{2}$O$_{4}$. Due to which the AFM contribution to $J_{1}$ suppress and FM contribution to $J_{1}$ wins for CdCr$_{2}$O$_{4}$ and HgCr$_{2}$O$_{4}$ for $U >$2 eV and $U \geq$2 eV, respectively.\cite{Yaresko} 
   
   From above discussion, it is clear that the inconsistency of $\varTheta$$_{CW}$ and $J_{1}$ especially for CdCr$_{2}$O$_{4}$ and HgCr$_{2}$O$_{4}$ with increasing $U$ may be due to the following two reasons. First may be due to the LSDA+$U$ calculations, which are often converged to local minima. Hence, it is expected that some out of many spin configurations considered by Yaresko may converge into the local minima for higher $U$ and hence leads to the inconsistency of $\varTheta$$_{CW}$ and $J_{1}$ for these two compounds. Second may be due to the unconstrained nature of calculations for various spin configurations as shown in our earlier publication in more details for ZnV$_{2}$O$_{4}$ compound. In that work, we have shown that the normal LSDA+$U$ calculations (the magnitude of magnetic moment (MM) of every V atoms is allowed to vary self-consistently for FM and AFM solutions) are not the correct method for predicting the experimentally reported AFM ground state of this compound for large parameter range of $U$. However, the constrained LSDA+$U$ calculations (the MM of every V atoms is fixed for both FM and AFM solutions) are able to predict the experimentally observed AFM ground state for large parameter range of $U$.\cite{Lal}
   
    Here, we try to understand the above issue related to the inconsistency about the nature and variation of magnitude of {\it J$_{1}$} ($\arrowvert${\it J$_{1}$}$\arrowvert$) as a function of $U$ in above mentioned compounds by using LSDA+$U$ method. The nature and the variation of $\arrowvert${\it J$_{1}$}$\arrowvert$ with increasing $U$ in unconstrained calculations are not consistent with the experimental data and not according to the relation, {\it J$_{1}$}$\propto$$\frac{t^{2}}{U}$ for CdCr$_{2}$O$_{4}$ and HgCr$_{2}$O$_{4}$ for $U >$3 eV and U=2-6 eV, respectively. In these calculations, a consistent behavior of {\it J$_{1}$} in our work is observed according to the experimental results and the above mentioned relation for ZnCr$_{2}$O$_{4}$ and MgCr$_{2}$O$_{4}$ for whole range of $U$. Such a behavior observed here for {\it J$_{1}$} is almost similar to Yaresko for ZnCr$_{2}$O$_{4}$, CdCr$_{2}$O$_{4}$ and HgCr$_{2}$O$_{4}$ compounds for $U$=2-4 eV. However, constrained calculations are found to give the nature and variation of $\arrowvert${\it J$_{1}$}$_{c}$$\arrowvert$ in the present study similar to that of experimental data and above mentioned relation for all four compounds. Among these compounds, magnitude of ($\varTheta$$_{CW}$)$_{c}$ is largest for ZnCr$_{2}$O$_{4}$ and smallest for HgCr$_{2}$O$_{4}$ for $U$=2-5 eV, which are according to experimentally reported order for these spinels.  
   
\section{Computational details}

    FM and AFM electronic-structure calculations of ACr$_{2}$O$_{4}$ (A=Zn, Cd, Mg and Hg) compounds are performed by using the {\it state-of-the-art} full-potential linearized augmented plane wave (FP-LAPW) method as implemented in elk code.\cite{elk} The experimentally observed structural parameters used in these calculations for every compounds in the face centered cubic phase are taken from the literature.\cite{Dutton,Chung2013,Wessels,Martin}  In these calculations, we have used Perdew -Wang/Ceperley -Alder exchange correlation functional.\cite{Perdew} The effect of on-site Coulomb interaction between Cr 3$d$ electrons is considered within the LSDA+$U$ approach.\cite{Bultmark} Normally in this method $U$ and $J$ are used as parameters. However, in present study, only $U$ is used as a free parameter and the value of $J$ is calculated self-consistently as described in reference [28]. In order to understand the magnetic behaviour of these compounds, we have done both unconstrained and constrained collinear magnetic calculations by varying $U$ from 0-6 eV, where the direction of MM is fixed along the z-axis. In unconstrained calculations, the magnitude of MM of every Cr atoms inside the muffin-tin sphere is allowed to vary self-consistently for both FM and AFM solutions. However, in the constrained calculations, the value of MM for every Cr atoms in FM solution is kept same to the self-consistently obtained value of MM in AFM solution. In this range of $U$, every calculations are started from the different combinations for converged electron densities and potentials corresponding to different values of $U$ for every compounds. In this manuscript, we have presented the data corresponding to $U$=2-6 eV. The atomic sphere radii are chosen to be 2.0, 2.4, 1.8, 2.5, 2.0 and 1.54 Bohr for Zn, Cd, Mg, Hg, Cr and O, respectively. 6x6x6 k-point grid has been used in the calculations. Convergence target of total energy has been set below ~10$^{-4}$ Hartrees/cell.

\section{Results and discussions}

    In order to know the nature of the nearest neighbour exchange coupling constant ({\it J$_{1}$}) and $\varTheta$$_{CW}$ for ACr$_{2}$O$_{4}$ (A=Zn, Cd, Mg and Hg) compounds, we have performed the unconstrained and constrained LSDA+$U$ calculations for FM and AFM spin structures. The FM and AFM face centered cubic primitive unit cell containing four Cr atoms (Cr1Cr2Cr3Cr4) with spin $\uparrow$$\uparrow$$\uparrow$$\uparrow$ and $\downarrow$$\uparrow$$\uparrow$$\downarrow$, respectively are sufficient to estimate the {\it J$_{1}$} and $\varTheta$$_{CW}$ for these compounds. The atomic and spin arrangements of Cr atoms in the AFM unit cell of these spinels are shown in the Fig. 1. Now, in order to reproduce the nature and the variation of magnitude of {\it J$_{1}$} reported by Yaresko with increasing $U$ in ACr$_{2}$O$_{4}$ (A=Zn, Cd and Hg) compounds, we first start with the unconstrained LSDA+$U$ calculations. In these calculations, the magnitude of MM of every Cr atoms is allowed to vary self-consistently for both FM and AFM solutions of these compounds. The plot of the total energy difference between FM and AFM solutions (${\Delta}E$=$E_{AFM}$-$E_{FM}$) per formula unit as a function of $U$ is shown in the Fig. 2. It is clear from the figure that for HgCr$_{2}$O$_{4}$, unconstrained calculations give the energy of FM solution less than the AFM solution when $U$ changes from 2-6 eV. However, the experimentally observed ground state for these compounds is AFM.\cite{Rudolf} Hence, unconstrained calculations fail to predict the AFM ground state for HgCr$_{2}$O$_{4}$ for whole range of $U$ studied here. For CdCr$_{2}$O$_{4}$, these calculations give the AFM ground state below $U$=4 eV because the total energy of AFM solution is less than FM solution. However, for $U \geq$4 eV, the total energy of $E_{FM}$$<$$E_{AFM}$ indicates that these calculations fail to predict the AFM ground state. For both ZnCr$_{2}$O$_{4}$ and MgCr$_{2}$O$_{4}$, unconstrained calculations give the AFM ground state for whole range of $U$ as the total energy of $E_{AFM}$$<$$E_{FM}$. The cause of the wrong ground state, especially for HgCr$_{2}$O$_{4}$ ($U$=2-6 eV) and CdCr$_{2}$O$_{4}$ ($U \geq$4 eV) compounds are expected because of the unconstrained nature of calculations.\cite{Lal} 
    
    We estimate the values of nearest neighbour exchange coupling constant {\it J$_{ij}$} among Cr atoms for these spinels by using the calculated energies of above mentioned FM and AFM spin configurations in the classical Heisenberg model of following formula,\cite{Ashcroft} 
     
     \begin{equation}
       {\it H}=-\sum_{\it i>j}{\it J_{ij}}S_{i}.S_{j} 
    \end{equation}
    
where, {\it S$_{i}$} and {\it S$_{j}$} are the Cr spins of sites i and j, respectively. In these spinels, the face centered cubic primitive unit cell contains six nearest neighbour interactions (where {\it J$_{ij}$}={\it J$_{1}$}) among magnetic Cr atoms. Hence, the total energy difference between FM and AFM spin ordering per unit cell in terms of {\it J$_{1}$} can be written as:

\begin{equation}
     {\it J_{1}}=\frac{{\it E_{AFM}}-{\it E_{FM}}}{18} 
    \end{equation}
    
for S=3/2 (for Cr 3$d^{3}$ electrons). 

   The calculated values of {\it J$_{1}$} for $U$=2-6 eV by using Eqn. (2) for these spinels are shown in the Table 1. It is evident from the table that for HgCr$_{2}$O$_{4}$, +ve values of {\it J$_{1}$} indicate the FM nature of the interaction for whole range of $U$. In this range of $U$, the values of {\it J$_{1}$} increases with the increase of $U$ are not according to the following equation,\cite{Saul}  
    
\begin{equation}
    {\it J}\approx\frac{4t^{2}}{U}.
\end{equation}

For CdCr$_{2}$O$_{4}$, the -ve values of {\it J$_{1}$} indicate the AFM nature of interaction for $U$=2-3 eV, where the magnitude of {\it J$_{1}$} ($\arrowvert${\it J$_{1}$}$\arrowvert$) decreases in accordance with the above equation. For $U >$3 eV, the nature of interaction becomes FM upto $U$=6 eV. The values of {\it J$_{1}$} increases continuously when $U$ changes from 4-6 eV as opposite to the Eqn. (3). For ZnCr$_{2}$O$_{4}$ and MgCr$_{2}$O$_{4}$, the AFM nature of the {\it J$_{1}$} is found for $U$=2-6 eV, where $\arrowvert${\it J$_{1}$}$\arrowvert$ decreases rapidly with the increase of $U$ as per expectation. Here, it is interesting to compare the calculated values of {\it J$_{1}$} in the present work with that estimated by Yaresko for ZnCr$_{2}$O$_{4}$, HgCr$_{2}$O$_{4}$ and CdCr$_{2}$O$_{4}$ compounds.\cite{Yaresko} The calculated values of {\it J$_{1}$} by Yaresko for these compounds are discussed in introduction in more details. The nature of {\it J$_{1}$} and the variation of $\arrowvert${\it J$_{1}$}$\arrowvert$ with increasing $U$ from 2-4 eV in the present work are similar to that of Yaresko for both ZnCr$_{2}$O$_{4}$ and HgCr$_{2}$O$_{4}$ compounds. In this range of $U$, such a behaviour of {\it J$_{1}$} in present work is also similar to that of Yaresko for CdCr$_{2}$O$_{4}$ except at $U$=3 eV. At this value of $U$, Yaresko has predicted {\it J$_{1}$} to be FM in nature as opposite to the present work. It is also clear from the table that the value of {\it J$_{1}$} estimated here is $\sim$2-3 times different from that calculated by Yaresko for some values of $U$ may be due to the following reason. It is well known that the LSDA+$U$ calculations are often converged to local minima. Hence, it is expected that some out of many spin configurations considered by Yaresko may converge into the local minima. However to avoid this factor, we have considered only two spin configurations (FM and AFM) for calculating {\it J$_{1}$} for these spinels. Also, all calculations for these two spin configurations are started from the different combinations for converged electron densities and potentials corresponding to different values of $U$ for every compounds. Hence, it appears that our calculations are not converged into the local minima. The inconsistent nature of {\it J$_{1}$} and the variation of $\arrowvert${\it J$_{1}$}$\arrowvert$ with increasing $U$ from 2-4 eV in the present and Yaresko work especially for CdCr$_{2}$O$_{4}$ and HgCr$_{2}$O$_{4}$ may be due to the unconstrained nature of calculations.

    Now, in order to verify whether the unconstrained calculations are really responsible for the inconsistent nature of {\it J$_{1}$} and the variation of $\arrowvert${\it J$_{1}$}$\arrowvert$ with increasing $U$ from 2-4 eV in the present and Yaresko work especially for CdCr$_{2}$O$_{4}$ and HgCr$_{2}$O$_{4}$, we have performed the constrained calculations for both spin configurations (FM and AFM) as a function of $U$. In constrained LSDA+$U$ calculations, we have fixed the magnitude of MM for every Cr atoms of FM solution similar to that of AFM one of above mentioned spinels. The plot of the total energy difference between FM and AFM solutions obtained from constrained calculations (${\Delta}E$$_{c}$=$E_{AFM}$-$E_{FM}$) per formula unit as a function of $U$ is shown in Fig. 3. It is evident from the figure that the constrained calculations give the AFM ground state for both HgCr$_{2}$O$_{4}$ and CdCr$_{2}$O$_{4}$ compounds for whole range of $U$ as the total energy of $E_{AFM}$$<$$E_{FM}$. For ZnCr$_{2}$O$_{4}$ and MgCr$_{2}$O$_{4}$, ${\Delta}E$$_{c}$$<$${\Delta}E$ suggests that the AFM ground state become more stable in constrained calculations for whole range of $U$. Hence the constrained calculations give the experimentally observed AFM ground state for HgCr$_{2}$O$_{4}$ and CdCr$_{2}$O$_{4}$ also for $U$=2-6 eV and $U \geq$4 eV, respectively where the unconstrained calculations failed. 
    
    The values of {\it J$_{1}$}$_{c}$ (where c stands for constrained calculation) are calculated by using Eqn. (2), where the total energies of FM and AFM per unit cell are used. The calculated values of {\it J$_{1}$}$_{c}$ for these compounds are also shown in the Table 1. It is clear from the table that the magnitude of {\it J$_{1}$}$_{c}$ ($\arrowvert${\it J$_{1}$}$_{c}$$\arrowvert$) decreases continuously for all four compounds as per expectation as according to Eqn. (3). The values of $\arrowvert${\it J$_{1}$}$_{c}$$\arrowvert$ for ZnCr$_{2}$O$_{4}$$>$MgCr$_{2}$O$_{4}$$>$CdCr$_{2}$O$_{4}$$>$HgCr$_{2}$O$_{4}$ when $U$ changes from 2 to 5 eV, which indicate that the strength of the magnetic interaction is largest in ZnCr$_{2}$O$_{4}$ and smallest in HgCr$_{2}$O$_{4}$. This is because of the different ionic size of A site Hg$^{2+}$, Cd$^{2+}$, Mg$^{2+}$ and Zn$^{2+}$ in ACr$_{2}$O$_{4}$ compounds. The ionic size of Hg$^{2+}$ is large as compared to Cd$^{2+}$, Mg$^{2+}$ and Zn$^{2+}$, which results the small values of {\it J$_{1}$}$_{c}$ for HgCr$_{2}$O$_{4}$ because of the large Cr-Cr distance as compared to other compounds. Above discussion clearly show that the inconsistency about the nature and the variation of $\arrowvert${\it J$_{1}$}$\arrowvert$ with increasing $U$ from 2-4 eV obtained in the unconstrained calculations especially for CdCr$_{2}$O$_{4}$ and HgCr$_{2}$O$_{4}$ have been removed by the constrained calculations.  
    
   Now, in order to compare the values of ($\varTheta$$_{CW}$)$_{c}$ with the ($\varTheta$$_{CW}$)$_{exp}$, we have estimated ($\varTheta$$_{CW}$)$_{c}$ by using the calculated values of {\it J$_{1}$}$_{c}$ for these compounds in the following formula,\cite{Ashcroft} 
    
\begin{equation}
   (\varTheta_{CW})_{c}=\frac{S(S+1)}{3k_{B}}z{\it J_{\rm 1}}_{c}  
    \end{equation}

where, $S$=3/2 for Cr 3$d^{3}$ electrons. $k$$_{B}$ and $z$=6 are the Boltzmann constant and the number of nearest neighbours of Cr atoms in the unit cell among which the exchange interaction are effective, respectively. The calculated values of ($\varTheta$$_{CW}$)$_{c}$ for these compounds for $U$=2-6 eV are shown in the Table 2. It is evident from the table that the absolute values of ($\varTheta$$_{CW}$)$_{c}$ [$\mid$($\varTheta$$_{CW}$)$_{c}$$\mid$] decreases from 735-87 K and 817-17 K for HgCr$_{2}$O$_{4}$ and CdCr$_{2}$O$_{4}$, respectively as $U$ varies from 2-6 eV. Similarly, for ZnCr$_{2}$O$_{4}$ and MgCr$_{2}$O$_{4}$, the values of $\mid$($\varTheta$$_{CW}$)$_{c}$$\mid$ decreases from 2034-834 K and 1895-539 K, respectively when $U$ changes from 2 to 6 eV. Now, we compare the calculated values of ($\varTheta$$_{CW}$)$_{c}$ with the experimental data. The experimentally reported values of ($\varTheta$$_{CW}$)$_{exp}$ for ZnCr$_{2}$O$_{4}$, MgCr$_{2}$O$_{4}$, CdCr$_{2}$O$_{4}$ and HgCr$_{2}$O$_{4}$ are ($\sim$-392 K, $\sim$-398 K), ($\sim$-350 K, $\sim$-346 K), ($\sim$-83 K, $\sim$-71 K) and $\sim$-32 K, respectively.\cite{Baltzer,Ueda2007,Rudolf} The values of $\mid$($\varTheta$$_{CW}$)$_{c}$$\mid$ for ZnCr$_{2}$O$_{4}$$>$MgCr$_{2}$O$_{4}$$>$CdCr$_{2}$O$_{4}$$>$HgCr$_{2}$O$_{4}$, which are consistent with the order of the magnitude of ($\varTheta$$_{CW}$)$_{exp}$ for $U$=2-5 eV. The calculated values of ($\varTheta$$_{CW}$)$_{c}$ for ZnCr$_{2}$O$_{4}$ and MgCr$_{2}$O$_{4}$ are $\sim$2.5 and $\sim$1.5 times larger than the experimental one, respectively somewhere around $U$=5 eV. Similarly, for HgCr$_{2}$O$_{4}$ and CdCr$_{2}$O$_{4}$, the values of ($\varTheta$$_{CW}$)$_{c}$ are $\sim$3.5 and $\sim$1.5 times larger than the experimental data, respectively. The large values of ($\varTheta$$_{CW}$)$_{c}$ in this work as compared to the experimental data is expected because of the following reasons: (i) the values of ($\varTheta$$_{CW}$)$_{exp}$ depends on which temperature range the reciprocal of magnetic susceptibility versus temperature data is fitted. For example, in some literatures ($\varTheta$$_{CW}$)$_{exp}$ were found to be -70 K, (-400 K, -433 K) and (-390 K, -400 K) for CdCr$_{2}$O$_{4}$, MgCr$_{2}$O$_{4}$ and ZnCr$_{2}$O$_{4}$, respectively and are different from the above mentioned values.\cite{Takagi,Dutton} and (ii) Solovyev $et$ $al$. have shown that the even in MnO, LSDA+$U$ method overestimates the magnetic interactions as compared to experimental results. This is due to the fact that this method does not treats the magnetic interactions properly and hence is expected to overestimate the values of ($\varTheta$$_{CW}$)$_{c}$ for these compounds.\cite{Solovyev} Hence, the calculated values of ($\varTheta$$_{CW}$)$_{c}$ in the present study are consistent with the experimental data as well as according to Eqn. (3) for large parameter range of $U$. Hence, the present work clearly suggests the importance of constrained calculations in these compounds.

\section{Conclusions}

    In conclusion, the inconsistency reported by [Yaresko, Phys. Rev. B. {\bf 77}, 115106 (2008)] in the theoretically estimated nature and the variation of magnitude of nearest neighbour exchange coupling constant ($\arrowvert${\it J$_{1}$}$\arrowvert$) as a function of $U$ in ACr$_{2}$O$_{4}$ (A=Zn, Cd, Mg and Hg) compounds have been studied by using LSDA+$U$ approach. The nature and the variation of magnitude of nearest neighbour exchange coupling constant ($\arrowvert${\it J$_{1}$}$\arrowvert$) in unconstrained calculations were not found to be consistent with the experimental data and the relation, {\it J$_{1}$}$\propto$$\frac{t^{2}}{U}$ for CdCr$_{2}$O$_{4}$ and HgCr$_{2}$O$_{4}$ for $U >$3 eV and U=2-6 eV, respectively. In this range of $U$, these calculations were found to give the behaviour of $\arrowvert${\it J$_{1}$}$\arrowvert$ according to  the experimental results and above mentioned relation for ZnCr$_{2}$O$_{4}$ and MgCr$_{2}$O$_{4}$. Such a behavior observed here for $\arrowvert${\it J$_{1}$}$\arrowvert$ were found to almost similar to that obtained by Yaresko for ZnCr$_{2}$O$_{4}$, CdCr$_{2}$O$_{4}$ and HgCr$_{2}$O$_{4}$ compounds as $U$ varies from 2 to 4 eV. However, the nature and variation of $\arrowvert${\it J$_{1}$}$_{c}$$\arrowvert$ observed during constrained calculations in the present work were found to be in accordance with the experimental data and above mentioned relation for all four compounds for large parameter range of $U$. Among these compounds, the values of $\arrowvert${\it J$_{1}$}$_{c}$$\arrowvert$ was largest for ZnCr$_{2}$O$_{4}$ and smallest for HgCr$_{2}$O$_{4}$, which indicate that the strength of magnetic interaction was largest for ZnCr$_{2}$O$_{4}$ and smallest for HgCr$_{2}$O$_{4}$ for $U$=2-5 eV. The order of magnitude of Curie-Weiss temperature [($\varTheta$$_{CW}$)$_{c}$] observed here was according to the experimentally reported order of magnitude of ($\varTheta$$_{CW}$)$_{exp}$ for these compounds. Hence, present study shows the importance of constrained calculations in understanding the magnetic behaviour of these complex systems.       

%\section{Acknowledgements}

\acknowledgments {S.L. is thankful to UGC, India, for financial support.}

\pagebreak 

\section{Tables}

TABLE I. Exchange coupling constants, {\it J$_{1}$} (meV) and {\it J$_{1}$}$_{c}$ (meV) [in bracket] obtained from total energy difference between AFM and FM spin ordering within the unconstrained and constrained LSDA+{\it U} calculations, respectively for ACr$_{2}$O$_{4}$ (A=Zn, Mg, Cd and Hg) compounds.

TABLE II. Curie-Weiss temperature, ($\varTheta$$_{CW}$)$_{c}$ (K) calculated from the nearest neighbour exchange coupling constant, {\it J$_{1}$}$_{c}$ obtained from the constrained LSDA+$U$ calculations for ACr$_{2}$O$_{4}$ (A=Zn, Mg, Cd and Hg) compounds.

\begin{table}[ht]
\caption{}
\begin{tabular}{p{0.9cm}p{2.8cm}p{2.8cm}p{3.2cm}p{3.2cm}}
\hline
{U}& HgCr$_{2}$O$_{4}$&CdCr$_{2}$O$_{4}$&ZnCr$_{2}$O$_{4}$&MgCr$_{2}$O$_{4}$ \\
(eV)&{\it J$_{1}$}({\it J$_{1}$}$_{c}$)&{\it J$_{1}$}({\it J$_{1}$}$_{c}$)&{\it J$_{1}$}({\it J$_{1}$}$_{c}$)&{\it J$_{1}$}({\it J$_{1}$}$_{c}$)\\
\hline
2&1.13(-8.46)&-1.40(-9.37)&-10.64(-23.36)&-10.74(-21.86)\\
3&1.22(-5.32)&-0.53(-6.74)&-8.06(-17.77)&-8.17(-13.69)\\
4&1.55(-3.32)&0.29(-4.57)&-6.08(-14.47)&-6.21(-12.44)\\
5&1.79(-1.41)&0.74(-1.71)&-4.50(-11.53)&-4.64(-8.36)\\
6&1.86(-1.01)&1.12(-0.20)&-3.28(-9.59)&-3.40(-6.23)\\
\hline
\end{tabular}
\end{table}

\begin{table}[ht]
\caption{}
\begin{tabular}{p{0.9cm}p{2.5cm}p{2.5cm}p{2.5cm}p{2.5cm}}
\hline
{U}& HgCr$_{2}$O$_{4}$&CdCr$_{2}$O$_{4}$&ZnCr$_{2}$O$_{4}$&MgCr$_{2}$O$_{4}$ \\
(eV)&($\varTheta$$_{CW}$)$_{c}$&($\varTheta$$_{CW}$)$_{c}$&($\varTheta$$_{CW}$)$_{c}$&($\varTheta$$_{CW}$)$_{c}$\\
\hline
2&-735&-817&-2034&-1895\\
3&-462&-582&-1547&-1182\\
4&-288&-391&-1260&-1078\\
5&-122&-147&-982&-721\\
6&-87&-17&-834&-539\\
\hline
\end{tabular}
\end{table}
\pagebreak
\section{Figure Captions:}

FIG. 1. The atomic and spin arrangements of all four Cr atoms in the antiferromagnetic face centered cubic primitive unit cell of ACr$_{2}$O$_{4}$ (A=Zn, Mg, Cd and Hg) compounds.

FIG. 2. The total energy difference between antiferromagnetic and ferromagnetic (${\Delta}E$=$E_{AFM}$-$E_{FM}$) solutions per formula unit obtained from the unconstrained LSDA+$U$ calculations as a function of $U$ for ACr$_{2}$O$_{4}$ (A=Zn, Mg, Cd and Hg) compounds.

FIG. 3. The total energy difference between antiferromagnetic and ferromagnetic (${\Delta}E$$_{c}$=$E_{AFM}$-$E_{FM}$) solutions per formula unit obtained from the constrained LSDA+$U$ calculations as a function of $U$ for ACr$_{2}$O$_{4}$ (A=Zn, Mg, Cd and Hg) compounds.

\begin{figure}
  \begin{center}
    \includegraphics{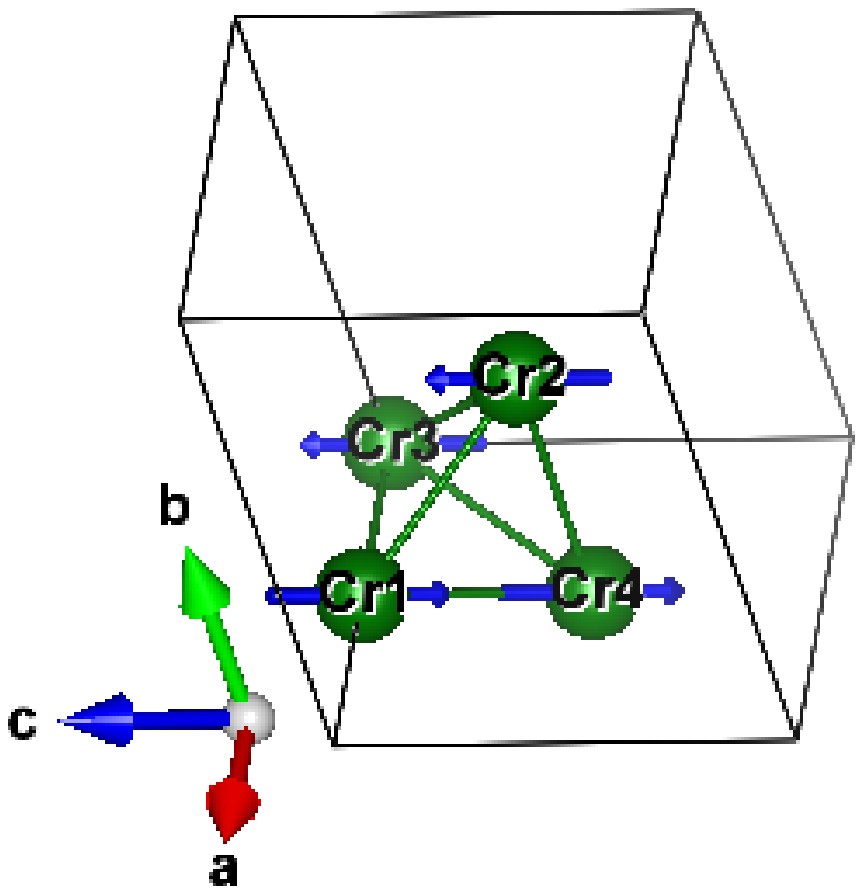}
    \label{}
    \caption{}
  \end{center}
\end{figure}

\begin{figure}
  \begin{center}
    \includegraphics{Fig2.eps}
    \label{}
    \caption{}
  \end{center}
\end{figure}

\begin{figure}
  \begin{center}
    \includegraphics{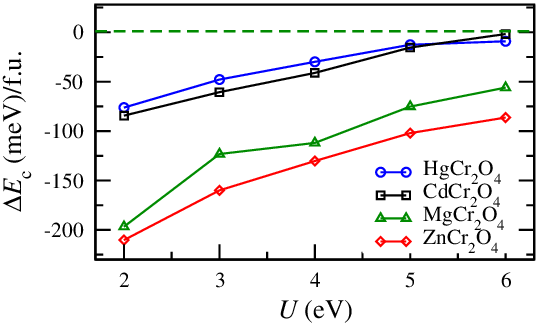}
    \label{}
    \caption{}
  \end{center}
\end{figure}

\end{document}